\documentclass[aps,floats,superscriptaddress,showpacs]{revtex4}
\usepackage{amsmath}

\usepackage{epsfig}
\usepackage{graphicx}
\usepackage{dcolumn}
\usepackage{bm}

\def\gapp{\lower.35em\hbox{$\stackrel{\textstyle>}{\sim}$}}
\def\lapp{\lower.35em\hbox{$\stackrel{\textstyle<}{\sim}$}}

\begin{document}
\bibliographystyle{apsrev}
%

\title{Minimal conductivity of rippled graphene with topological disorder}

\author{Alberto Cortijo and Mar\'{\i}a A. H. Vozmediano}
\affiliation{Unidad Asociada CSIC-UC3M,
Instituto de Ciencia de Materiales de Madrid,\\
CSIC, Cantoblanco, E-28049 Madrid, Spain.}

\date{\today}
\begin{abstract}
We study the transport properties of a neutral graphene sheet with
curved regions induced or stabilized by topological defects. The
proposed model gives rise to  Dirac fermions in a random magnetic
field and in the random space dependent Fermi velocity  induced by
the curvature. This last term leads to singular long range
correlated disorder with special characteristics. The Drude
minimal conductivity at zero energy is found to be inversely
proportional to the density of topological disorder, a signature
of diffusive behavior.

\end{abstract}
%
%
%
%

\maketitle
 \section{Introduction}
 Since its experimental realization  \cite{Netal04,Netal05,ZTSK05},
 graphene has been a focus of intense
research activity both theoretical and experimentally.  The origin
of this interest lies partially on the experimental capability of
explore the transport properties that show a set of interesting
features related to disorder and which open the way to graphene
based electronics \cite{GN07}.

One of the  most intriguing properties of graphene is the
observation of a minimal conductivity at zero frequency in undoped
suspended  samples that in early measurements was argued to have a
universal value of the order of $e^{2}/h$
\cite{Netal05,Netal06,KNG06} independent of the disorder
concentration and a factor of $\pi$ bigger than the one predicted
by
theory\cite{Fradkin2,L93,MH97,PGC06,K06,K06b,OGM06,AE06,A06,Z06}.
Other experiments in both mechanically deposited and graphene
grown on a substrate \cite{ZTSK05,ZSAK05,Yetal07} have found a
bigger dispersion in the coefficient of the universal behavior
while more recent calculations \cite{Z07,HAS07,AHGS07} have casted
some doubts on the disorder dependence of the numerical
coefficient and the actual situation remains unclear.

Another peculiarity of most of the graphene samples is the
existence of mesoscopic corrugations
\cite{Metal07,Setal07,Ietal07} whose possible influence on
transport properties only now starts to be explored
\cite{Metal07,Metal07b,K07b,KG07,HJV07}. Although the observed
ripples were invoked from the very beginning to explain the
absence of weak localization in the samples
\cite{Metal06,MG06,KNG06}, there have been so far few attempts to
model the corrugations based either in the curved space approach
with \cite{CV07a,CV07b} or without topological defects
\cite{JCV07}, or on the theory of elasticity
\cite{Metal07,FLK07,CK07,ANetal07}. The possible physical
implications of the ripples in connection with the charge
anisotropies has been revisited in very recent works
\cite{GKV07,BP07}.

In refs. \cite{CV07a,CV07b} we proposed a model for rippled
graphene based on the presence defective rings (pentagons and
heptagons) in the  samples. These types of topological defects
have been observed in nanotubes and in bombarded graphite and are
known to be a natural way to get rid of tensions in the hexagonal
lattice \cite{CJV07}. It is then natural to think that some of the
defects that were either present in the graphite sample or formed
during the very energetic procedure of mechanical cleavage, stay
quenched in the two dimensional samples.

It is clear by now that the low energy electronic properties of
graphene are very well described by the massless Dirac equation in
two dimensions, a fact coming from the symmetries of the hexagonal
lattice and also obtained in the tight binding approximation
\cite{W47}. It is also known that the Dirac points are very robust
to deformations of the lattice \cite{MGV07} so we have modelled
curved graphene  assuming that the Dirac points are not affected
by the presence of ripples and hence that the curved samples can
be described by writing the Dirac equation in the given curved
surface \cite{BD82}. Within this formalism studied recently the
electronic structure of the sample with topological defects
\cite{CV07a,CV07b} and that of graphene with smooth curved regions
\cite{JCV07}. In the last work we emphasized the fact that
curvature gives rise not only to an effective magnetic field, a
property known from the early times of graphene
\cite{GGV92,GGV93,GGV93a}, but also to an effective position
dependent Fermi velocity which can have strong influence on the
physical properties of the system.

In this work we  continue studying the physical properties of
curved graphene  within the "gravity" approach. The electronic
properties were explored by means of the two point Green's
function of the electron and we could make advances keeping a
fixed number of defects at given positions of the lattice. To
study the transport properties is a much more difficult task. We
need to assume a density of defects with some statistical
distribution and to average over defects. We apply the standard
techniques of disordered electrons \cite{AS06} to rippled graphene
by averaging over the random magnetic fields and over the
effective Fermi velocity induced by curvature. We will make
special emphasis on the case of having topological defects. The
smooth curvature case can be implemented easily. We find that
averaging over the space-dependent Fermi velocity treated as a
random scalar field in the case of having  topological defects,
gives rise to a singular, long range correlated interaction that
affects severely the one particle properties of the system. We
compute semiclassically the zero frequency conductivity and find
that it  depends on the inverse of the density of disorder, a
behavior characteristic of diffusive systems. We argue that even
if the Drude value obtained in this work gets renormalized by
quantum corrections to the universal minimal conductivity there
will still be a region in parameter space where this model differs
from the ones studied previously.

The article is organized as follows: In Section \ref{model} we
review  the model of topological defects and establish the
effective Hamiltonian. In section \ref{averaging}   we define the
statistical properties of the fields induced by the defects and
apply the replica trick to get the effective four Fermi
interaction. We see that the interaction is anisotropic and
singular in the forward direction, a behavior  due to the long
range character of the conical singularities. We discuss this
issue and extract the interaction coefficients that will appear in
the computation of the lifetime and the density of states of the
disordered system. Section \ref{sigmamodel} contains the
semiclassical expansion of the sigma model to get the low energy
behavior of the system. We discuss two possible saddle points and
compute the effective potential  to establish the symmetry
breaking minimum, a basic  point to the rest of the calculation.
From there we deduce, after some lengthly calculations detailed in
the Appendix A, the diffusive behavior of the system. Section
\ref{discussion} contains a discussion of the results and open
questions.

\section{The model}
\label{model}
In this section we follow closely refs
\cite{CV07a,CV07b} to describe the explicit form of the potential
generated by the defects and its statistical properties. The model
for corrugated graphene is based on the presence of defective
carbon rings at arbitrary positions in the lattice. It is known
that substitution of an hexagon by an n-sided polygon with n
greater (smaller) than six gives rise to locally curved portions
in the sample with positive (negative) curvature. It was argued
that the presence of an equal number of heptagons and pentagons
would keep the sample flat in average and mimic the corrugations
observed in the free standing samples. In the mentioned
references, we studied the electronic properties of the sample by
computing the electron Green's function in the curved space
defined by a given number of heptagons and pentagons located at
fixed positions of the lattice. In order to study transport
properties we need to consider a density of defects located at
random positions.

The behavior of the electrons in curved graphene is described by
the Hamiltonian:
\begin{equation}
H=iv_{F}\int d^2\mathbf{r}\sqrt{g}\bar{\psi}\gamma^\mu
\left(\partial_{\mu}-\Omega_{\mu}(\mathbf{r})\right)\psi.
\label{curvedaction}
\end{equation}
The curved Dirac matrices $\gamma^\mu({\bf r})$ are related with
the usual constant matrices $\gamma^a$ by
$$\gamma^\mu({\bf r})=e^{\mu}_{a}(\mathbf{r})\gamma^a,$$ where
$e^{\mu}_{a}$ is the tetrad constructed with the metric tensor. It
is this factor in combination with the determinant of the metric
$\sqrt{g}$ gives rise to a space dependent effective Fermi
velocity.  In terms of a tight binding language this term can be
modelled as a global modulation of the nearest neighbor hopping
\cite{GKV07} or of the average distance between carbon atoms
\cite{BP07} induced by curvature. The term $\Omega_\mu$ contains
the spin connection and the possible extra gauge fields induced by
the defects \cite{CJV07}. It is given by
\begin{equation}
\Omega_\mu({\bf r})=\Gamma_{\mu}(\mathbf{r})-
\tau_{\alpha}A^{\alpha}_{\mu}(\mathbf{r})
\label{gaugef}
\end{equation}
where $\Gamma_\mu$ is due to the spin connection and the
non-abelian part $\tau_{\alpha}A^{\alpha}_{\mu}$ is related to the
holonomy  and will be discussed later.

In \cite{CV07a,CV07b} we described the curved space generated by a
arbitrary number of topological defects by the  metric
\begin{equation}
g_{ij}=e^{\Lambda(\mathbf{r})}\delta_{ij}
\end{equation}
 where the conformal
factor $\Lambda(\mathbf{r})$ takes the form
\begin{equation}
\Lambda(\mathbf{r})=\sum_{j}^{N}\frac{\mu_{j}}{2\pi}\log\left|\frac{r}{a^{*}}\right|,
\label{conformalfactor}
\end{equation}
where $\mu_{j}$ is a constant related to the defect (or excess)
angle of the disclinations  and $a^{*}$ is a constant of the order
of the lattice spacing, interpreted as the radius of the "core" of
the defect.  The specific form of the conformal factor
(\ref{conformalfactor}) gave rise to strongly diverging one
particle properties like the local density of states. We will see
that  in the procedure of averaging over disorder one particle
properties will remain singular while two particle properties,
like the Drude conductivity, will be convergent.

It is well known that the presence of topological defects has
other consequences besides of curving the graphene sheet. Odd
membered rings can mix the Fermi points that can be implemented by
a nonabelian gauge field \cite{GGV92,GGV93}. Also if various
defects are present, an extra phase appears due to the non
commutativity of the holonomy operators associated to the valley
mixing phase and the proper Berry phase that fermions get when
they surround the group of defects \cite{OK01,LC04}. All these
phases are naturally incorporated in this formalism by generic
external nonabelian gauge fields $A_{\alpha}(\mathbf{r})$ in
equation (\ref{curvedaction}). This also accounts for the
disordered graphene classification described in  \cite{AE06,SA06}.
In the case of having an equal number of  pentagon and heptagonal
defects it can be shown that only the (abelian) gauge field
associated to the conical singularity remains \cite{CJV07} we can
restrict ourselves to the scattering problem around a single Fermi
point. We will comment on the possible modifications of the
calculation that a more general case would induce in section
\ref{discussion}.

The value of $|\mu_{j}|\equiv\mu$ is $1/24$ for both pentagon and
heptagon rings. We use $\mu$ as a perturbative parameter around
flat space and expand the determinant of the metric
$g(\mathbf{r})$, the zweibeins $e^{\mu}_{a}(\mathbf{r})$ and the
spin connection $\Omega(\mathbf{r})$ in (\ref{curvedaction}). To
first order in $\mu$ the Hamiltonian is
\begin{widetext}
\begin{eqnarray}
H=iv_{F}\int d^{2}\mathbf{r}[(1+\Lambda({\bf
r}))\bar{\psi}\gamma^{i}\partial_{i}\psi+
\frac{i}{2}\bar{\psi}\gamma^{i}(\partial_i\Lambda({\bf r}))\psi].
\label{defectaction}
\end{eqnarray}
\end{widetext}
Eqs.  (\ref{conformalfactor}) and (\ref{defectaction}) are the
basis of our calculation. We have a disorder-induced modification
of the Fermi velocity described by $\Lambda(\mathbf{r})$ and a
gauge field-type term if we identify
$A_{i}\sim\partial_{i}\Lambda(\mathbf{r})$. Notice that although
the gauge potential $A_{i}$ seems to be a total derivative, the
effective magnetic field is not zero due to the singular form of
the function $\Lambda(\mathbf{r})$ (\ref{conformalfactor}). Also
due to the presence of the conformal factor $\Lambda(\mathbf{r})$
in the effective Fermi velocity of the Hamiltonian
(\ref{defectaction}) even if  the potential could be be gauged
away by means of a singular gauge transformation (a coordinate
choice in our case), its effects on the observable quantities will
show up as coming from the other term.

\section{Averaging over disorder}
\label{averaging}
We will study the transport properties of this
topologically disordered graphene by assuming a random
distribution of an equal number of pentagons and heptagons. The
statistical properties of these topological defects where analyzed
in part in \cite{GGV01}. It was proposed there that the gaussian
disorder induced by a random distribution of topological defects
can be described by a single dimensionless quantity $\Delta$
proportional to the average fluctuations of the -non abelian-
vector potential representing a vortex at the position of the
defect.
\begin{equation}
\langle {\bf A} ( {\bf r} ) {\bf A} ( {\bf r'} ) \rangle = \Delta
\delta^2 ( {\bf r} - {\bf r'} ) \label{delta}
\end{equation}
The infrared behavior of   $\Delta$ was analyzed in \cite{GGV01}
and shown to diverge logarithmically with the size $R$ of the
system in the case of unbounded disclinations
\begin{equation}
\Delta = 2 \pi \Phi_0^2 \log \left( \frac{R}{l_0} \right)
\label{vortex}
\end{equation}
while it remains a of constant value
\begin{equation}
\Delta \propto \Phi_0^2 n_{disl} b^2 \label{disloc}
\end{equation}
in the case of having pentagon-heptagon pairs bounded into
dislocations with density  $n_{disl}$ and with an average distance
$b$.

We will work in the intermediate case of having a random
distribution of an equal number of five and seven rings at average
distances $d$ such that the "volume" occupied by  virtual strings
pairing the defects is small compared with the total size of the
sample $(d/L)^2\ll 1$.

The most important issue in this work is the new random field
$\Lambda(\mathbf{r})$ associated to the Fermi velocity
modification given in eq. (\ref{conformalfactor}).  We will assume
for this scalar field a zero mean value
$\left<\Lambda(\mathbf{r})\right>=0$ and
%
\begin{equation}
\left<\Lambda(\mathbf{r})\Lambda(\mathbf{r}')\right>=na^{*2}
\log\left|\frac{\mathbf{r}-\mathbf{r}'}{a^{*}}\right|,
\label{statcorrelator}
\end{equation}
where $n$ is proportional to the areal density of defects and
$a^{*}$ is of the order of the lattice spacing. This correlator
diverges in both the infrared and the ultraviolet limits. It
induces over the random magnetic field $A_i({\bf r})$ the average
\begin{equation}
\left<A_i({\bf r})A_j({\bf r'})\right>\sim
na^{*2}\delta_{ij}\delta(\mathbf{r}-\mathbf{r}').
\end{equation}

The origin of the correlator (\ref{statcorrelator}) can be
understood when we consider the nature of the defects. In a
geometrical description of defects in  two dimensional crystals
\cite{SN88}, the  equations of motion for the metric tensor
$g_{ij}(\mathbf{r})$ reduce to a unique equation for the conformal
factor (\ref{conformalfactor}):
\begin{equation}
\nabla^{2}\Lambda(\mathbf{r})=\sum_{j}^{N}
\frac{\mu_{j}}{2\pi}\delta(\mathbf{r}-\mathbf{r}_{j}).
\label{poisson equation}
\end{equation}
We can rewrite (\ref{conformalfactor}) in terms of the Green's
function of  (\ref{poisson equation}):
\begin{equation}
\Lambda(\mathbf{r})=\sum_{j=1}^{N}\frac{\mu_{j}}{2\pi}\int
d\mathbf{r}'\delta(\mathbf{r}'-
\mathbf{r}_{j})K(\mathbf{r}-\mathbf{r}'). \label{green equation}
\end{equation}
where the Green's function has de form
\begin{equation}
K(\mathbf{r}-\mathbf{r}')\approx\log\left|\frac{\mathbf{r}-\mathbf{r}'}{a^{*}}\right|,
\label{greensfunction}
\end{equation}
for $\mathbf{r}\gg a^{*}$, what justifies equation
(\ref{statcorrelator}).

In momentum space the correlators are
\begin{equation}
\left<\Lambda(\mathbf{p})\Lambda(\mathbf{-p})\right>=\frac{na^{*2}}{p^2}
\label{corrlambdap}
\end{equation}
\begin{equation}
\left<A_i(\mathbf{p})A_j(\mathbf{-p})\right>=na^{*2}\delta_{ij}
\end{equation}
Notice that the $\Lambda$ term describes  a new type of disorder:
Dirac Fermions in two space dimensions with a random velocity, a
problem that, to our knowledge has not been addressed in the early
literature \cite{Letal94,CMW96,NTW94,HD02}.

In momentum space representation the action corresponding to eq.
(\ref{defectaction}) reads:
\begin{equation}
S=\frac{1}{2}\int
dk\bar{\psi}\left(\omega\gamma^0-v_F\vec{\gamma}\vec{k}\right)\psi
-\frac{i}{2} v_F\int dk
\frac{1}{2}\bar{\psi}\vec{\gamma}\vec{A}(\vec{k})\psi - \frac{
v_F}{2}\int dp dk \Lambda(p) \bar{\psi}\vec{\gamma}\vec{k}\psi,
\label{disorder_action}
\end{equation}
where $dk\equiv\frac{d^{2}\mathbf{k}}{4\pi^{2}}$ and
\begin{equation}
\Lambda({\bf p})=\int\frac{d^2
x}{2\pi}e^{ipx}\log\vert\frac{x}{a}\vert.
\end{equation}
We integrate out $\Lambda$ using eq. (\ref{corrlambdap}) as a
quadratic action.  It is interesting to note that this term is
similar to the interaction between curvatures in a continuous
hexatic membrane where $1/na^{*2}$ plays the role of the hexatic
stiffness constant \cite{SN88}.

Replicating the fields and integrating out $\Lambda$ in
(\ref{disorder_action}) we get:
\begin{widetext}
\begin{equation}
S=S_{0}+v_{F}^{2}\frac{\lambda}{2}\int dk\;dk'\Gamma(k, k')
(\bar{\psi}_{a}\gamma^{i}\psi_{a})
(\bar{\psi}_{b}\gamma_{i}\psi_{b}), \label{effaction}
\end{equation}
\end{widetext}
where summation over replica indices $a,b$ is assumed,
$\lambda=2\pi\mu^2 n a^2$ is a dimensionless parameter
proportional to the density of defects $n$ and
\begin{equation}
\Gamma(k, k')=\left(
\frac{(k+k')^{2}}{(k-k')^{2}}-\frac{1}{4}\right).
\label{Gamma}
\end{equation}
The constant term in the interaction vertex (\ref{Gamma}) comes
from the random magnetic field and would give rise to the standard
result found in previous works. The term coming from the Fermi
velocity is anisotropic and singular when $k\to k'$, a signature
of the infrared singularity associated to the effect of a conical
defect at infinite distances from the apex.
 Because we are dealing with elastic scattering, the
modulus of the momentum is conserved provided that the energy
$\omega$ in the process is conserved. The function  $\Gamma(k,
k')$ in (\ref{Gamma}) can be written as a function of the
difference of the scattering angles,
$\phi\equiv\theta_{k}-\theta_{k'}$:
\begin{equation}
\Gamma(\phi)=\left(
\frac{\cos^{2}(\phi/2)}{\sin^{2}(\phi/2)}-\frac{1}{4}\right).
\label{Gamma2}
\end{equation}
This function diverges at scattering angles $\phi=0$ (forward
scattering) and the one particle properties of the system like the
density of states or the imaginary part of the self-energy will
show anomalous behavior when compared with their counterparts in
short ranged scattering processes. The divergence in
(\ref{Gamma2}) can be  regularized by including a cutoff $\delta$:
\begin{equation}
\Gamma(\phi,\delta)=\left(
\frac{\cos^{2}(\phi/2)}{\sin^{2}(\phi/2)+\delta^{2}}-\frac{1}{4}\right).
\label{Gamma3}
\end{equation}
The meaning of the cutoff can be understood from the origin of the
correlator for the function $\Lambda(\bf r)$ in eq.
(\ref{statcorrelator}). Instead of having a long ranged propagator
corresponding to infinite range defects that behave like
$\frac{1}{p^2}$, we may consider a correlator of the type
\begin{equation}
K({\bf p})\sim\frac{1}{p^2+\delta^2} \label{kcondelta}
\end{equation}
what corresponds in real space  to changing (\ref{greensfunction})
by a modified Bessel function of the second kind:
\begin{equation}
K(\bf r -\bf r')=K_{0}(\delta\left|\bf r -\bf r'\right|).
\label{bessel}
\end{equation}
The leading term in the expansion of (\ref{bessel}) for small
$\delta$ is
\begin{equation}
K(\bf r -\bf r')\approx\log(\delta\left|\bf r -\bf r'\right|).
\end{equation}
Using eq. (\ref{kcondelta}) instead of eq. (\ref{corrlambdap}) we
arrive at eq. (\ref{Gamma3}), where we have redefined $\delta$ as
$\delta=\delta/k^2$. We can assume that the momenta involved in
the problem are of the order of the wavelength $\lambda$ of the
states near the Dirac points (or the localization length  if those
states are localized) so we can consider that $\delta$ is of the
order of $\lambda/\chi$ where $\chi$ is the biggest length scale
of the problem comparable to the system size $L$. We will make the
important assumption that $\chi>\lambda$. We will see that the
Drude conductivity is independent of $\delta$ and then well
defined.

 The free action in (\ref{effaction}) is the usual:
\begin{equation}
S_{0}=\int dk\bar{\psi}_{a}(\omega\gamma^{0}-v_{F}\gamma k +i\eta
M)\psi_{a}, \label{freeaction}
\end{equation}
The replica indices run over $1$ to $2N$. The first $N$ indices
are associated to advanced fields ($+i\eta$ in the free action),
and the second ones to retarded fields, related to $-i\eta$, with
the obvious definition of the matrix $M$:
\begin{equation}
M=\left(
          \begin{array}{cc}
            1& 0 \\
            0 & -1 \\
          \end{array}
        \right)_{2N}.
\label{lambda}
\end{equation}
This $\eta$-term breaks explicitly the $O(2N)$ symmetry of the
action (\ref{effaction}) leading to possible massless excitations
\cite{MS81}.

In addition to the forward divergence the scattering function
(\ref{Gamma3}) is highly anisotropic \cite{WB84} what will lead us
to deal  with a multichannel scattering problem in contrast with
the sort ranged scattering case where the scattering only occurs
in the s-channel. We will see that  despite the anisotropic
scattering the diffusion process will have an isotropic behavior
described by a scalar diffusion constant.

In order to keep track of the role of each channel, we decompose
$\Gamma(\phi)$ in harmonics:
\begin{equation}
\Gamma(\phi)=\sum_{n}\Gamma_{n}e^{-i n\phi}.
\label{Gamma_harm}
\end{equation}
We will see later that only the $n=0,\pm1$ channels will play a
role (notice  that $\Gamma_{n}=\Gamma_{-n}$). Their explicit
values as a function of the cutoff $\delta$ are:
\begin{equation}
\Gamma_{0}=\frac{1}{\delta}-5/4+O(\delta)\;\;,\;\;
\Gamma_{1}=\frac{1}{\delta}-2+O(\delta).
\label{coeff}
\end{equation}

\section{The saddle point approximation in the nonlinear $\sigma$ model and results}
\label{sigmamodel}
In this section we will  construct the
low-energy field theory describing the large scale behavior of the
system (i.e., at length scales larger than the elastic mean free
path, $l$) following standard procedures \cite{AS06}.

The interaction term in (\ref{effaction}) can be written as
\begin{widetext}
\begin{equation}
S_{int}=\frac{1}{2}\lambda v_{F}^{2}\sum_{n}\int (dk)(dk')
\Gamma_{n}\chi_{n}(\hat{k})\chi_{n}^{*}(\hat{k'})(\bar{\psi}_{a}\gamma^{i}\psi_{a})
(\bar{\psi}_{b}\gamma_{i}\psi_{b}), \label{channelaction}
\end{equation}
\end{widetext}
where $\chi_{n}(\hat{k})=e^{in\theta _{k}}$. Now we proceed to
simplify the model (\ref{channelaction}). The current-current
interaction in (\ref{channelaction}) can be transformed into a
density-density type term \cite{CMT87}:
\begin{widetext}
\begin{equation}
S_{int}=-\frac{1}{2}gv_F^2\sum_{n}\int dk dk'
\Gamma_{n}\chi_{n}(\hat{k})\chi_{n}^{*}(\hat{k'})[(\bar{\psi}_{a}\psi_{b})
(\bar{\psi}_{b}\psi_{a})- (\bar{\psi}_{a}\gamma_5\psi_{b})
(\bar{\psi}_{b}\gamma_5\psi_{a})],
 \label{channelaction2}
\end{equation}
\end{widetext}
where we have redefined the coupling constant as
$g\equiv-2\lambda$.

The  four fermi interaction in (\ref{channelaction2}) can be
decoupled by means of a Hubbard-Stratonovitch trnasformation
\begin{equation}
S=S_{0}+\int dk \frac{1}{4\pi
gv^2_{F}}[\sum_{n}\Gamma_{n}\left(Q^{2}_{n}+\Pi^{2}_{n}\right)+i
\sum_{n}\Gamma_{n}\chi_{n}(\hat{k})\bar{\psi}\left(Q_n-
i\gamma_{5}\Pi_n\right)\psi], \label{hubbard1}
\end{equation}
where $Q_{n}$ and $\Pi_{n}$ are Hubbard-Stratonovitch fields.
Because we have defined retarded and advanced fermionic fields,
$Q$ and $\Pi$ will be $2N$ dimensional matrix fields.

We can further simplify the calculations by performing a unitary
transformation that diagonalizes the fermionic part of the action
in (\ref{hubbard1}), and leaves unchanged the functional
integration measure: $\psi\rightarrow U\psi$,
$\bar{\psi}\rightarrow \hat{\psi}U^{+}$, $Q_{n}\rightarrow
UQ_{n}U^{+}$, $\Pi_{n}\rightarrow U\Pi_{n}U^{+}$,
$\gamma_{5}\rightarrow U\gamma_{5}U^{+}$. Without loss of
generality we will name those transformed fields as the original
ones.
Integrating out the fermionic modes leads to the usual final form
for the non linear $\sigma$ model:
\begin{equation}
S_{eff}=\int Tr\ln G^{-1}-\frac{1}{4
gv^2_{F}}\sum_{n}\Gamma_{n}(Q^2_{n}+\Pi^{2}_{n}),
\label{finalaction}
\end{equation}
where the Green's function is defined by
\begin{equation}
G^{-1}=G^{-1}_{0}+i\eta M
+i\sum_{n}\Gamma_{n}\chi_{n}(\hat{k})(Q_{n}-i\gamma_{5}\Pi_{n}),
\label{greenfunction}
\end{equation}
and
\begin{equation}
G^{-1}_{0}=\left(\begin{array}{cc}
                   w+v_{F}k & 0 \\
                   0 & w-v_{F}k
                 \end{array}
\right)\otimes \mathbf{1}_{2N}.
\label{diaggreen}
\end{equation}

Next we will make a saddle point approximation  and seek for a
solution of the saddle point equation:
\begin{equation}
\frac{\delta S_{eff}}{\delta \left<S_{n}\right>}=0,
\label{saddle1}
\end{equation}
In contrast with the case of isotropic short ranged scatterers
\cite{Fradkin1,Fradkin2} where  a single equation is obtained,
(\ref{saddle1}) represents an infinite number of coupled
equations:
\begin{equation}
\frac{1}{2gv^{2}_{F}}\left<S_{n}\right>=\int
(dk)\frac{\chi_{n}(\hat{k})}{G_{0}^{-1}+
i\sum_{n}\Gamma_{n}\chi_{n}(\hat{k})\left<S_{n}\right>}.
\label{saddle2}
\end{equation}
Following \cite{AWM94}, in the limit $\omega\rightarrow 0$, and
despite of the fact that the scattering mechanism is anisotropic,
we can try to find a solution of the type
\begin{equation}
\left<Q_n \right> =f \delta_{n0}M.
\label{spsolution}
\end{equation}
Of course, the solution of the mean field equations is not unique,
and other vacua with different properties may be found. The value
of the trial function $f$  is:
\begin{equation}
f=\frac{v_{F}K}{\Gamma_{0}} \left(e^{\frac{2\pi}{g\Gamma_{0}}}-1
\right)^{-\frac{1}{2}},
\label{brokensolution}
\end{equation}
where $K$ is an ultraviolet cutoff. Now we are free to make a
choice over the mean field values of $Q$ and $\Pi$. The standard
choice is $\langle Q_{0}\rangle=\langle S_{0}\rangle$ and
$\langle\Pi_{0}\rangle=0$. With this solution for $\left<Q\right>$
we arrive to the one particle relaxation time or lifetime
\begin{equation}
\frac{1}{2\tau}\equiv v_{F}K \left(e^{\frac{2\pi}{g\Gamma_{0}}}-1
\right)^{-\frac{1}{2}},
\label{lifetime}
\end{equation}
which turns out to be a constant whose dependence on the disorder
strength is  typically non-perturbative: $\tau^{-1}(g)\sim
\exp(1/g)$ . This value for the lifetime depends on both the
ultraviolet cutoff $K$, and on the infrared cutoff $\delta$
through the scattering coefficient
$\Gamma_{0}\sim\frac{1}{\delta}-5/4$.
The
ultraviolet cutoff W can be removed by renormalization group
techniques as done in \cite{Fradkin1} but the infrared cutoff
$\delta$ remains and the one-particle properties of the theory
will depend explicitly on it \cite{KY03}. We will see later that
the transport properties ( i.e. two-particle properties) are
independent of such regulator \footnote{For a more extended
discussion about this issue in the context of usual two
dimensional electron gas see the refs.
\cite{AWM94},\cite{AI92}).}.
\\

We can now compute the averaged density of states at the Fermi
level,
\begin{equation}
\rho(0)=-\frac{1}{N\pi}Im \int\frac{d^2k}{4\pi^2}G^{R}(k),
\end{equation}
the index $N$ in the denominator is the replica index and the
limit $N\rightarrow 0$ will be  taken at the end of the
calculation. From now on, we drop any reference to this limit,
bearing in mind that it must be taken when computing observable
quantities.

The density of states as a function of the lifetime
(\ref{lifetime}) is:
\begin{equation}
\rho(0)=\frac{1}{g \Gamma_{0}v^{2}_{F}}\frac{1}{2\tau}.
\label{DOS}
\end{equation}
%

Before proceeding to compute the quantum fluctuations around the
chiral symmetry breaking solution of the saddle point equations we
will make a comment on the solution $\langle S_{n}\rangle=0$. This
solution leads to a vanishing density of states at the Fermi
energy and to a sublinear frequency dependence:
$\rho(\omega)\sim|\omega|^{\alpha}$, with $\alpha$ being a
function of the strength of the disorder, a behavior reported in
\cite{Letal94}. In order to determine the true minimum of the $Q$
field action we have computed the effective potential as a
function of $f$ defined as in (\ref{brokensolution}). This
effective potential is the same as that of the Nambu-Jona Lasinio
model \cite{T58,NJL61,FI01} being a function of $f$ defined as in
(\ref{brokensolution}), when $\langle\Pi_{0}\rangle$ is taken to
be zero:
\begin{equation}
V_{eff}=\frac{1}{4gv^{2}_{F}}f^2+\frac{f^2}{4\pi}\left(\log
\left(\frac{f^2}{K^2}\right)-1\right).\label{effectivepotencial}
\end{equation}
The  result  is well known.  Minimizing this potential the two
possible solutions are again the solution $f=0$ corresponding to a
zero value for the effective potential and
 the symmetry breaking solution $f=Ke^{1/2-K^{2}/2}e^{\frac{-\pi}{2g}}$
 giving $V_{eff}=-\frac{K^2}{4\pi}f^2$. This broken symmetry solution
 equivalent to (\ref{brokensolution})  is then a minimum of the theory.

 We will then proceed computing the physical properties of the
 quantum field model built around the broken symmetry solution.
 We will see that the physics obtained for this case
 is typically non-perturbative irrespective of the strength of the
 disorder.

The technical details of the rest of the computation are given in
appendix A. Once we have calculated the value of the leading
configuration of $Q_{n}$ from the saddle point equations
(\ref{saddle1}), we expand the action (\ref{finalaction}) around
this value, setting $Q_{n}=\left<Q_{n}\right>+\delta Q_{n}$ and
retain in the expansion terms up to second order in $\delta
Q_{n}$:
%
\begin{equation}
S\approx\left<S\right>+\delta Q_{n}\frac{\delta^{2}S^{*}}{\delta
Q_{n}\delta Q_{m}}\delta Q_{m}+... \label{actionexpansion}
\end{equation}
%
The $^{*}$  means that the functional derivative is evaluated in
the saddle point solution for $Q_{n}$ and $\Pi_{n}$.

The ultimate goal  is to compute the action for the massless
modes:
\begin{equation}
\delta S=\int dq\delta Q_{0}\frac{(2\tau)\Gamma_{0}}{4
gv^2_F}\left(\eta+Dq^2\right)\delta Q_{0}, \label{ultimate_action}
\end{equation}
from where we can extract  the diffusion constant is $D$. From eq.
(\ref{ultimate_action})  we get
\begin{equation}
D=\frac{1}{2\pi^{2}}\frac{1}{g\rho(0)}\frac{1}{\Gamma_{0}-\Gamma_{1}},
\label{diffconstant}
\end{equation}
where the coefficients $\Gamma_{0}$ and $\Gamma_{1}$ are given in
(\ref{coeff}) and their difference is
$\Gamma_{0}-\Gamma_{1}=\frac{3}{4}+O(\delta)$, hence the diffusion
coefficient is well defined when the cutoff $\delta$ is send to
zero.

Finally we can compute the semiclassical value for the static $DC$
conductivity using the Einstein relation for the two diffusive
channels $\delta Q_{0}$ and $\delta\Pi_{0}$ in eq.
(\ref{channelaction2}) :
\begin{equation}
\sigma_{DC}=4\frac{e^2}{\hbar}\rho(0)D=\frac{4e^2}{h}2\frac{4}{3\pi
g} \label{dccond}
\end{equation}
The factor of 4  comes from the spin and valley degeneracy and the
factor 2 comes from the two diffusive channels.

The result in (\ref{dccond})  depends on the coupling parameter
$g=4\pi\mu^2 n a^2$, which contains information about the type of
disorder  ($\mu$) and the density of disorder $n$.

\section{Discussion and open questions}
\label{discussion}
 In this work we have addressed the effects of
curvature on the transport properties of corrugated graphene
sheets. We have shown that coupling the Dirac field to a curved
space gives rise to an effective potential whose general form and
statistical properties depend on the metric. The main feature of
the geometrical description is the appearance of an effective
space dependent Fermi velocity which gives rise to a random scalar
field coupled to the kinetic energy term in the Hamiltonian.

Smooth curved regions in graphene give rise to standard short
range correlated disorder as the one studied in the literature
\cite{Fradkin2}. The presence of  topological defects in the
sample either as the main source of curvature or as a way to
stabilize the ripples in the mechanically deposited samples gives
rise to singular, long range correlated disorder that leads the
system to an intrinsically  diffusive behavior even at the
neutrality point and without the need for external scattering
sources.

We have studied the behavior of the conductivity at zero voltage
of a graphene sample in presence of a finite density of
topological defects. The result given in eq. (\ref{dccond}) shows
that the conductivity is proportional to the inverse of the
density of defects, a distinctive feature of a diffusive system. A
similar non universal behavior has been found in the same system
and has been attributed to the effects of random coulomb
scatterers present in the substrate
\cite{NM06,HAS07,AHGS07,Tetal07}. A crucial difference is that in
the mentioned references  the graphene sample is either heavily
doped or it has a nonzero carrier density  due to a local field
effect induced by the Coulomb impurities. In our work the density
of states is generated by the disorder as in refs.
\cite{PGC06,Fradkin2}.

Another noticeable feature  of the model presented in this work is
the strong dependence of the one particle properties on the
parameter $\delta$ regulating the infrared behavior of the model.
The situation here is even worse than that of a two dimensional
electron gas  in a long range correlated random magnetic field
discussed in ref. \cite{AWM94}. There  the one particle relaxation
time was found to diverge as the infrared cutoff is sent to zero
but the finite density of states made the transport relaxation
time $\tau_{tr}$  finite. In our case $\tau_{tr}$ also depends on
the density of states at the Fermi level but now the DOS is
divergent for $\delta\rightarrow 0$. As we have seen despite this
singular behavior we have obtained   a finite Drude conductivity
due to the particular dependence of the diffusive constant with
the density of states (eq. (\ref{diffconstant})).

In  section III we have introduced the parameter $\delta$ defined
in a phenomenological fashion as the ratio between the
characteristic  length scale  of the defect $\chi$, and the wave
or localization length $\lambda$ of the states around the Fermi
energy. In a semiclassical  approximation to the problem this
parameter is essentially  uncontrollable. We can nevertheless made
an estimation of the range of applicability of our results by
assuming that the localization length can be obtained from an
analysis of the quantum corrections to the conductivity. The
diffusive regime is characterized by a static mean free path
$l=v_{F}\tau_{tr}$ greater than the localization or wavelength
$\lambda$ but smaller than the system's size. The mean free path
can be estimated using expressions (\ref{lifetime}), (\ref{DOS}),
and (\ref{diffconstant}) and assuming that $a^{*}\sim a$ and $\chi
\sim L$.  We thus find
\begin{equation}
l \sim \frac{2}{3\mu}\left(
\frac{\chi}{\lambda}\right)^{1/2}\frac{1}{n_{imp}^{1/2}}.
\end{equation}
from where we can get a  lower bound for the density of defects in
the case $\lambda<l$:
\begin{equation}
n_{imp}<\frac{4}{9\mu^2}\frac{L}{\lambda^3}.
\end{equation}
In the same spirit, an upper bound can be estimated using the
condition $l<L$:
\begin{equation}
n_{imp}>\frac{4}{9\mu^2}\frac{1}{L\lambda}.
\end{equation}
In refs. \cite{ASZ02,OGM06} the possible fixed points of the total
conductivity where classified according to the symmetries of the
original Hamiltonian in a renormalization group analysis. As our
disorder term preserves both chiral and time reversal symmetries,
the  final conductivity once quantum corrections are taken into
account should flow to the universal value of $4e^2/\pi h$. We
note that in previous works  this universal value is already
obtained at the Drude level. The topological disorder discussed in
this work sets as an initial condition of the RG flow a rather
different - disorder dependent - value that can - or not - flow to
the usual fixed point. The analysis of the quantum corrections to
the conductivity (\ref{dccond}) is beyond the scope of this work
and will be reported somewhere else.

\section{Acknowledgments}
We gratefully acknowledge useful discussions with K. Ziegler, D.
Khveshchenko and F. Guinea. We also thank  E. Fradkin, P.
W\"{o}lfle and I. V. Gornyi for kindly explaining their work to
us. This work was supported  by  MEC (Spain) through grant
FIS2005-05478-C02-01 and by the European Union Contract 12881
(NEST).

\appendix
\section{Sigma model calculations}

Once we have calculated the value of the leading configuration of
$Q_{n}$ from the saddle point equations (\ref{saddle1}), we expand
the action (\ref{finalaction}) around this value, setting
$Q_{n}=\left<Q_{n}\right>+\delta Q_{n}$ and retain in the
expansion terms up to second order in $\delta Q_{n}$:
%
\begin{equation}
S\approx\left<S\right>+\delta Q_{n}\frac{\delta^{2}S^{*}}{\delta
Q_{n}\delta Q_{m}}\delta Q_{m}+... \label{actionexpansion}
\end{equation}
%
The finite density of states of eq. (\ref{DOS}) allows us to use
the usual identity in the integration of momentum  derived for the
two dimensional electron gas: $$\int\frac{d^2
k}{4\pi^2}\rightarrow \rho(0)\int d\varepsilon_k \;\;,\;\;  v_F
k\rightarrow\varepsilon_k .$$ As a consistency test,  computing
the density of states with this change we get the expression
(\ref{DOS}). This change of variables will simplify  the
calculations. The $^{*}$ in (\ref{actionexpansion}) means that the
functional derivative is evaluated at the saddle point solution
for $Q_{n}$ and $\Pi_{n}$. Also, other terms of quadratic order
appear in (\ref{actionexpansion}) with crossing functional
derivatives in the fields $\delta Q$ and $\delta \Pi$. We will see
shortly that these derivatives are zero and the former fields are
not coupled.
 These derivatives are
\begin{equation}
\delta Q_{n}\frac{\delta^{2}S^{*}}{\delta Q_{n}\delta Q_{m}}\delta
Q_{m}=-\frac{1}{2}\sum_{n}\Gamma_{n}\delta
Q^{2}_{n}+\frac{1}{4}\int(dp)(dq)\sum_{n,m}\Gamma_{n}\Gamma_{m}\chi_{p}(k)\chi_{m}(p+q)\delta
Q_{n}TrG(k)G(p+q)\delta Q_{m},\label{derivative1}
\end{equation}
and
\begin{equation}
\delta \Pi_{n}\frac{\delta^{2}S^{*}}{\delta \Pi_{n}\delta
\Pi_{m}}\delta \Pi_{m}=-\frac{1}{2}\sum_{n}\Gamma_{n}\delta
\Pi^{2}_{n}-\frac{1}{4}\int(dp)(dq)\sum_{n,m}\Gamma_{n}\Gamma_{m}\chi_{n}(p)\chi_{m}(p+q)\delta
\Pi_{n}Tr\gamma_{5}G(p)\gamma_{5}G(p+q)\delta
\Pi_{m}.\label{derivative2}
\end{equation}

Since the spectral functions are peaked at the Fermi energy, we
can effectively restrict the $\mathbf{q}$  integration  to values
around the Fermi point, $q<<K_{F}$ and write
$\chi_{m}(p+q)\approx\chi_{m}(p)$ causing an error of the order of
$O(q)$. In (\ref{derivative1}) the integral will be dominated by
the product $G^{R}G^{A}$, i.e., by the off diagonal sector of the
fields $\delta Q^{+-}$ leading to the diffusive pole behavior for
the fields $Q_{n}$. By contrast, in (\ref{derivative2}) using the
symmetry property $\gamma_{5}G^{R}\gamma_{5}=-G^{A}$ we can see
that the diffusive pole will come from the product $G^{A}G^{A}$
and its retarded-retarded counterpart $\delta\Pi^{++,--}$ will be
the channel for the diffusive behavior (also, the minus sign
appearing in this symmetry corrects the relative sign between the
second terms in (\ref{derivative1}) and (\ref{derivative2})).

Let us define the quantity $C_{mn}(\eta,q)$ as
\begin{equation}
C_{mn}(\eta,q)=\frac{1}{4}\int(dp)\chi_{n}(p)\chi_{m}(p)TrG^{R}(p)G^{A}(p+q).
\label{diffuson1}
\end{equation}
To obtain an action for the modes $Q_{n}$ and $\Pi_{n}$ at small
$q$ and $\eta$, we expand $G^{R}$ up to first order in $\eta$ and
to second order in $q$ in equation (\ref{diffuson1}).

The first term in this expansion , without any reference to the
replica index, is
\begin{equation}
C_{nm}^{0}=\frac{1}{4}\int(dp)e^{i(n+m)\theta}Tr
                                                        \left(\begin{array}{cc}
                                                          \frac{1}{v_{F}p-i\epsilon} & 0 \\
                                                          0 & \frac{1}{-v_{F}p-i\epsilon} \\
                                                        \end{array}\right)
                                                        \left(\begin{array}{cc}
                                                          \frac{1}{v_{F}p+i\epsilon} & 0 \\
                                                          0 & \frac{1}{-v_{F}p+i\epsilon} \\
                                                        \end{array}\right).\nonumber
\end{equation}
We have denoted $\epsilon=\eta+\frac{1}{2\tau}$. In terms of these
variables, we have
\begin{equation}
C_{nm}^{0}=\frac{1}{4}\rho(0)\delta_{-nm}\frac{1}{\eta+\frac{1}{2\tau}}.
\label{zeroterm1}
\end{equation}
Expanding (\ref{zeroterm1}) up to first order in $\eta$, and using
(\ref{DOS}), we get
\begin{equation}
C_{nm}^{0}=\frac{\delta_{-nm}}{4
gv^{2}_{F}\Gamma_{0}}-\frac{1}{4}\rho(0)\delta_{-nm}(2\tau)^{2}\eta+O(\eta^{2}).
\label{zeroterm2}
\end{equation}
The constant term in (\ref{zeroterm2}) coincides with the term
proportional to $\delta Q_{n}^2$ and $\delta\Pi^{2}_{n}$ in
(\ref{derivative1}) and (\ref{derivative2}) respectively. This
mass contribution to the action is
\begin{equation}
\mathcal{L}_{m}\equiv\frac{1}{4
gv^{2}_{F}}\sum_n\Gamma_n\left(\frac{\Gamma_n}{\Gamma_0}-1\right)
\left(\delta Q_n^2+\Pi^2_n\right).
\end{equation}
We immediately see that only the modes $\delta Q_{0}$ and $\delta
\Pi_{0}$ are massless, and they will responsible for the diffusive
behavior of the system exactly as happens in the 2DEG
\cite{AWM94}. In what follows we will eliminate  the $\Pi$ field.
It represents another diffusion channel that does not mix with the
$Q$'s and plays the same role. We will simply multiply by two the
final result.
 The next term in the $q$ expansion reads
\begin{widetext}
\begin{equation}
C^{1}_{nm}=\frac{1}{4}\int(dp)e^{i(m+n)\theta}v_{F}q\cos\theta
\left(\frac{-1}{(v_{F}p-i\epsilon)(v_{F}p+i\epsilon)^2}+
\frac{1}{(-v_{F}p-i\epsilon)(-v_{F}p+i\epsilon)^2}\right).
\label{firstterm1}
\end{equation}
\end{widetext}
Note that in the case of  short ranged isotropic scattering this
term vanishes due to the angular integration. In our case,
however, the presence of $e^{i(m+n)\theta}$ allows linear terms in
$q$, coupling the massive modes $\delta Q_{\pm1}$  to $\delta
Q_{0}$. The integral in (\ref{firstterm1}) is performed changing
to the energy variable and noticing that the angular integration
gives a non-zero result only when $n=-m\pm1$:
\begin{equation}
C^{1}_{nm}=\frac{1}{4}v_{F}q(\delta_{nm-1}+\delta_{nm+1})\rho(0)\int
d\varepsilon_{p}\frac{1}{(\varepsilon_{p}-i\epsilon)(\varepsilon_{p}+i\epsilon)^{2}},
\end{equation}
or, after setting $\eta=0$,
\begin{equation}
C^{1}_{nm}=-\frac{v_{F}q}{4}\rho(0)(\delta_{nm-1}+\delta_{nm+1})(2\tau)^{2}.
\label{firstterm2}
\end{equation}
To compute the next term in the $q$ expansion, we will use the
following trick. The trace of the product of Green functions in
(\ref{diffuson1}) can be explicitly written as
\begin{equation}
TrG^{R}G^{A}=\left(\begin{array}{cc}
                       \frac{1}{v_{F}p-i\epsilon} & 0 \\
                       0 & \frac{1}{-v_{F}p-i\epsilon} \\
                     \end{array}
\right)\left(
         \begin{array}{cc}
           \frac{1}{v_{F}\left|\mathbf{p+q}\right|+i\epsilon} & 0 \\
           0 & \frac{1}{-v_{F}\left|\mathbf{p+q}\right|+i\epsilon} \\
         \end{array}
       \right),
\end{equation}
or, rearranging signs,
\begin{equation}
TrG^{R}G^{A}=
\frac{1}{v_{F}p-i\epsilon}\frac{1}{v_{F}\left|\mathbf{p+q}\right|+
i\epsilon}+\frac{1}{v_{F}p+i\epsilon}\frac{1}{v_{F}\left|\mathbf{p+q}\right|-
i\epsilon}.
\end{equation}
We immediately see that the second term in the right hand side is
the complex conjugate of the first term, thus,
\begin{equation}
TrG^{R}G^{A}=2Re\left(\frac{1}{v_{F}p-i\epsilon}
\frac{1}{v_{F}\left|\mathbf{p+q}\right|+i\epsilon}\right).
\label{trick1}
\end{equation}
The terms already calculated in the expansion of $C_{nm}(\eta,q)$
can be easily derived with this trick, but where we make a real
profit of this simplification is in the calculation of the term
$q^{2}$:
\begin{equation}
C^{2}_{nm}=\frac{1}{2}q^{2}Re
\int(dp)e^{i(m+n)\theta}\frac{1}{v_{F}p-i\epsilon}\left(
\frac{\cos^{2}\theta}{(v_{F}p+i\epsilon)^{3}}-
\frac{\sin^{2}\theta}{2v_{F}p(v_{F}p+i\epsilon)^{2}}\right).
\label{secondterm}
\end{equation}
If we compare the angular part of (\ref{secondterm}) with the
corresponding part in (\ref{firstterm1}) we see that after
performing the integral   in (\ref{secondterm}) there are terms of
the type $\delta_{-nm\pm2}$ together with terms $\delta_{-nm}$
which generate couplings between the zero modes $\delta Q_{0}$ and
the massive $\delta Q_{\pm2}$, and $\delta Q_{0}\delta Q_{0}$
respectively. The couplings involving $\delta Q_{\pm2}$ being of
order $q^{2}$ will produce  terms of order $q^{4}$ and can be
neglected.  We will only keep the terms independent of $\theta$ in
(\ref{secondterm}), from which we will extract the diffusion
coefficient $D$ for the massless diffusive mode $\delta Q_{0}$.
The result for (\ref{secondterm}) only taking into account the
terms proportional to $\delta_{-nm}$ is (we shift the pole at
$v_{F}p=0$ in the second term in the integrand and take the real
part, the first term will not contribute to this real part):
\begin{equation}
C^{2}_{nm}=\frac{v_{F}^{2}q^{2}}{8\pi}\delta_{-nm}\rho(0)(2\tau)^{3}.
\label{secondterm2}
\end{equation}
Collecting all the terms, the action for the modes $\delta Q_{0}$
and $\delta Q_{\pm1}$ is:
\begin{eqnarray}
\delta S_{Q}\approx&&\int
(dq)\frac{1}{4gv^{2}_{F}}\sum_{n\pm1}\Gamma_{n}\left(\frac{\Gamma_{n}}{\Gamma_{0}}-1\right)\delta
Q_{n}^{2}-\frac{\rho(0)(2\tau)^{2}}{4}\Gamma^{2}_{0}\delta
Q_{0}\left(\eta+\frac{v^{2}_{F}(2\tau)q^{2}}{2\pi}\right)\delta
Q_{0}\nonumber\\
&&+\frac{\Gamma_{1}\Gamma_{0}\rho(0)v_{F}q(2\tau)^2}{4}\left(\delta
Q_{0}\delta Q_{1}+\delta Q_{0}\delta
Q_{-1}\right).
\label{action_coupling}
\end{eqnarray}
In order to get a theory for the $n=0$ modes, we integrate out the
$n=\pm1$ modes in (\ref{action_coupling}). Using again (\ref{DOS})
we get
\begin{eqnarray}
\delta S=\int d q\frac{\rho(0)(2\tau)^2\Gamma^{2}_{0}}{4}\delta
Q_{0}\left(\eta+\frac{v^{2}_{F}q^{2}(2\tau)}{2\pi}+
\frac{\Gamma_{1}}{\Gamma_{0}-\Gamma_{1}}\frac{v^{2}_{F}q^{2}(2\tau)}{2\pi}\right)\delta
Q_{0}.\label{zeromode_action}
\end{eqnarray}
If we simplify and use again (\ref{DOS}) we arrive to the final
action for the massless modes:
\begin{equation}
\delta S=\int dq\delta Q_{0}\frac{(2\tau)\Gamma_{0}}{4
gv^2_F}\left(\eta+Dq^2\right)\delta Q_{0},
\label{ultimate_action}
\end{equation}

\bibliography{longrange}

\end{document}